\documentstyle[12pt]{article}

\font\twlgot =eufm10 scaled \magstep1
\font\egtgot =eufm8
\font\sevgot =eufm7

\font\twlmsb =msbm10 scaled \magstep1
\font\egtmsb =msbm8
\font\sevmsb =msbm7

\newfam\gotfam

\textfont\gotfam\twlgot
\scriptfont\gotfam\egtgot
\scriptscriptfont\gotfam\sevgot

\newfam\msbfam
\textfont\msbfam\twlmsb
\scriptfont\msbfam\egtmsb
\scriptscriptfont\msbfam\sevmsb
\def\Bbb{\protect\pBbb}
\def\pBbb{\relax\ifmmode\expandafter\Bb\else\typeout{You cann't use
Bbb in text mode}\fi}
\def\Bb #1{{\fam\msbfam\relax#1}}

\def\thebibliography#1{\bigskip\section*{\large
\bf References\\}\list
  {[\arabic{enumi}]}{\settowidth\labelwidth{#1}\leftmargin\labelwidth
    \advance\leftmargin\labelsep
    \usecounter{enumi}}
    \def\newblock{\hskip .11em plus .33em minus .07em}
    \sloppy\clubpenalty4000\widowpenalty4000
    \sfcode`\.=1000\relax}

\def\op#1{\mathop{{\it\fam0} #1}\limits}

\newcommand{\Ker}{{\rm Ker\,}}

\newcommand{\lng}{\langle}
\newcommand{\rng}{\rangle}

\newcommand{\beq}{\begin{equation}}
\newcommand{\eeq}{\end{equation}}
\newcommand{\ben}{\begin{eqnarray}}
\newcommand{\een}{\end{eqnarray}}
\newcommand{\be}{\begin{eqnarray*}}
\newcommand{\ee}{\end{eqnarray*}}
\newcommand{\bea}{\begin{eqalph}}
\newcommand{\eea}{\end{eqalph}}

\newcommand{\cA}{{\cal A}}

\newcommand{\cT}{{\cal T}}

\newcommand{\cE}{{\cal E}}
\newcommand{\cH}{{\cal H}}

\newcommand{\cD}{{\cal D}}

\newcommand{\bL}{{\bf L}}
\newcommand{\bH}{{\bf H}}

\newcommand{\f}{\phi}

\newcommand{\Om}{\Omega}

\newcommand{\g}{\gamma}
\newcommand{\G}{\Gamma}

\newcommand{\w}{\wedge}

\newcommand{\wt}{\widetilde}
\newcommand{\wh}{\widehat}
\newcommand{\ol}{\overline}

\newcommand{\dr}{\partial}

\newcommand{\ap}{\approx}

\newenvironment{eqalph}{\stepcounter{equation}
\setcounter{equationa}{\value{equation}}
\setcounter{equation}{0}

\begin{eqnarray}}{\end{eqnarray}\setcounter{equation}{\value{equationa}}}

\newcommand{\mar}[1]{}

\begin{document}
\hbox{}

{\parindent=0pt

{\large \bf Noether conservation laws in quantum mechanics}
\bigskip 

{\bf G. Sardanashvily}

\medskip

\begin{small}

Department of Theoretical Physics, Moscow State University, 117234
Moscow, Russia

E-mail: sard@grav.phys.msu.su

URL: http://webcenter.ru/$\sim$sardan/
\bigskip

{\bf Abstract.}
Being quantized, conserved Noether symmetry functions are represented
by Hermitian operators in the space of solutions of the Schr\"odinger
equation, and their mean values are conserved.
\end{small}
}

\bigskip
\bigskip

Classical non-relativistic time-dependent mechanics can be described as
a particular field theory on a fibre bundle $Q\to \Bbb R$ 
over the time axis $\Bbb
R$ \cite{book98,jmp00,sard98}. Its configuration space $Q$ is
equipped with bundle coordinates 
$(t,q^i)$, where
$t$ is the Cartesian coordinate on $\Bbb R$ possessing
transition functions $t'=t+$const. A fibre bundle $Q\to\Bbb R$ is trivial,
but its
different trivializations correspond to different non-relativistic
reference frames. 

Noether conservation laws in Hamiltonian mechanics issue 
from the invariance of an integral invariant of
Poincar\'e--Cartan  
under one-parameter groups of bundle isomorphisms of a
configuration space $Q\to\Bbb R$ \cite{epr}. Therefore, being quantized,
conserved Noether symmetry functions commute with the Schr\"odinger operator.
They act in the space of solutions of the Schr\"odinger
equation, and their mean values are conserved.

\section{Noether conservation laws in classical Hamiltonian mechanics}

The momentum phase space of non-relativistic
mechanics is the vertical
cotangent bundle $V^*Q$ of $Q\to\Bbb R$ equipped with holonomic coordinates 
$(t,q^i,p_i)$ \cite{book98,jmp00,sard98}. 
The cotangent
bundle $T^*Q$ of $Q\to\Bbb R$ coordinated by $(t,q^i,p,p_i)$ plays a role
of the homogeneous momentum phase space. 
It is provided with the canonical Liouville form $\Xi=pdt +p_idq^i$
and the canonical symplectic form $\Om=d\Xi$.
The corresponding Poisson bracket  reads
\mar{n90}\beq
\{f,f'\}_T=\dr^pf\dr_tf' +\dr^if\dr_if'-
\dr_tf\dr^pf' -\dr_if\dr^if', \quad \dr^p=\dr/\dr p, \quad
f,f'\in C^\infty(T^*Q). \label{n90}
\eeq
There is the trivial affine bundle 
\mar{n9}\beq
\zeta: T^*Q \to V^*Q. \label{n9}
\eeq
Due to this fibration, the vertical cotangent bundle $V^*Q$ is provided
with the canonical Poisson bracket
\mar{n91}\beq
\{f,f'\}_V=\dr^if\dr^if' -\dr_if\dr^if', \qquad f,f'\in C^\infty(V^*Q),
\label{n91}
\eeq
such that
\mar{n92}\beq
\zeta^*\{f,f'\}_V=\{\zeta^*f,\zeta^*f'\}_T, \label{n92}
\eeq
where $\zeta^*f$ denotes the pull-back onto $T^*Q$ of a function $f$ on
$V^*Q$.

A Hamiltonian of non-relativistic time-dependent mechanics is defined
as a section 
\mar{n10}\beq
h: V^*Q\to T^*Q, \qquad p\circ h=-\cH(t,q^j,p_j) \label{n10}
\eeq
of the fibre bundle (\ref{n9}).
The pull-back 
$h^*\Xi$ onto $V^*Q$ of the Liouville form $\Xi$ by means of a section $h$ 
(\ref{n10}) is the
well-known  integral invariant of Poincar\'e--Cartan
\mar{n11}\beq
H=p_idq^i-\cH dt. \label{n11}
\eeq
We agree to call it a Hamiltonian form. 
There exists a unique vector field $\g_H$ on $V^*Q$
 such that
\mar{n12}\ben
&& dt\rfloor \g_H=1, \qquad \g_H\rfloor dH=0, \nonumber\\
&& \g_H=\dr_t+\dr^i\cH\dr_i-\dr_i\cH\dr^i. \label{n12}
\een
It defines the first
order Hamilton equation 
\mar{n13}\beq
d_t q^i=\dr^i\cH, \qquad d_tp_i=-\dr_i\cH \label{n13}
\eeq
on $V^*Q$, where $d_t=\dr_t +q^i_t\dr_i + p_{ti}\dr^i$ is the
total derivative written with respect to the adapted
coordinates $(t,q^i,p_i,q_t^i,p_{ti})$ on 
the jet manifold $J^1V^*Q$ of the fibre bundle $V^*Q\to\Bbb R$.
Accordingly, a smooth real function $f$ on $V^*Q$ 
is an integral of motion if its Lie derivative
\mar{n93}\beq
\bL_{\g_H}f=\g_H\rfloor f=(\dr_t+\dr^i\cH\dr_i-\dr_i\cH\dr^i)f
\label{n93}
\eeq
along $\g_H$ vanishes.

In an equivalent way, let us consider
the pull-back $\zeta^*H$ of the Hamiltonian form
$H$ (\ref{n11}) onto $T^*Q$, and let us define the function
\mar{n70}\beq
\bH=\dr_t\rfloor (\Xi-\zeta^*H)=p+\cH \label{n70}
\eeq
on $T^*X$. 
Then, the relation
\mar{n72}\beq
\zeta^*(\bL_{\g_H}f)=\{\bH,\zeta^*f\}_T \label{n72}
\eeq
holds for any smooth real function $f$ on $V^*Q$. In particular,
$f$ is an integral of motion iff the bracket $\{\bH,\zeta^*f\}_T$
vanishes. We agree to call $\bH$ (\ref{n70}) the homogeneous Hamiltonian.

For the sake of simplicity, we will further denote
the pull-back $\zeta^*f$ onto $T^*Q$ of a function $f$ on $V^*Q$ by the
same symbol $f$, and will identify the Poisson algebra $C^\infty(V^*Q)$
with the Poisson subalgebra $\zeta^*C^\infty(V^*Q)$ of
$C^\infty(T^*Q)$.

A Noether conservation law in Hamiltonian mechanics issue from the
invariance of a Hamiltonian form $H$ (\ref{n11}) under a one-parameter
groups of bundle automorphisms of a configuration space $Q\to\Bbb R$
\cite{book98,jmp00,epr}. Its infinitesimal generator is a projectable
vector field 
\mar{n2}\beq
u=u^t\dr_t +u^i(t,q^j)\dr_i \label{n2}
\eeq
on $Q\to \Bbb R$, where $u^t=0,1$ because time reparametrizations are
not considered.   
If $u^t=0$,
we have 
a vertical vector field $u=u^i\dr_i$ which takes its values into the
vertical tangent bundle $VQ$ of $Q\to\Bbb R$. If $u^t=1$, a vector field 
$u$ (\ref{n2}) is a connection on the configuration 
bundle $Q\to\Bbb R$.
Note that connections 
\mar{n1}\beq
\G=\dr_t +\G^i(t,q^j)\dr_i \label{n1}
\eeq
on $Q\to\Bbb R$ make up an affine space
modelled over the vector space of vertical vector fields on $Q\to\Bbb
R$, i.e., the sum of a connection and a vertical vector field is a
connection, while the difference of two connections is a vertical
vector field on $Q\to\Bbb R$.   

Any projectable vector field $u$ (\ref{n2}) on 
$Q\to\Bbb R$ admits the canonical lift 
\mar{n8}\beq
\wt u=u^t\dr_t +u^i\dr_i-p_j\dr_i u^j\dr^i \label{n8}
\eeq
onto $V^*Q$. It generates a one-parameter group of holonomic bundle
automorphisms of the momentum phase space $V^*Q\to\Bbb R$. The
Hamiltonian form $H$ (\ref{n11}) is invariant under this group of
automorphisms iff its Lie derivative 
\mar{n42}\beq
\bL_{\wt u}H=\wt u\rfloor dH+ d(\wt u\rfloor H)=(\dr_t(p_iu^i-u^t\cH) -
u^i\dr_i\cH +p_j\dr_i u^j \dr^i\cH)dt
\label{n42}
\eeq
along $\wt u$ (\ref{n8}) vanishes. There is equality
\mar{n100}\beq
\bL_{\wt u}H= -\g_H\rfloor d\cT_u, \label{n100}
\eeq
where
\mar{n50}\beq
\cT_u=u^t\cH-u^ip_i \label{n50}
\eeq
is called the Noether symmetry function associated to a vector
field $u$. If the Lie derivative $\bL_{\wt u}H$ vanishes, then
\mar{n102}\beq
\g_H\rfloor d\cT_u=\{\bH,\cT_u\}_T=0, \label{n102}
\eeq
and the symmetry function 
$\cT_u$ (\ref{n50}) is an integral of motion. One can treat the equality
as the Noether conservation law $d_t\cT_u\ap 0$ on the shell 
(\ref{n13}) \cite{book98,epr}. 

Since $u^t=0,1$, there are the following two types of
Noether symmetry functions (\ref{n50}).
If $u=v=v^i\dr_i$ is a vertical vector field on $Q\to \Bbb R$,
then the symmetry function
\mar{n52}\beq
-\cT_v=v^ip_i \label{n52}
\eeq
is the momentum along $v$. 

If $u=\G=\dr_t +\G^i\dr_i$ is a connection, the corresponding symmetry
function  
\mar{n53}\beq
\cT_\G=\cH-p_i\G^i. \label{n53}
\eeq
is an energy function. However, it need not be a true physical energy.
There are different energy functions $\cT_\G$
(\ref{n53}) corresponding to different connections $\G$ on $Q\to\Bbb R$.
Moreover, if an energy function $\cT_\G$ is an integral of motion and
the momentum  
$\cT_v$ (\ref{n52}) is so, the energy function 
$\cT_{\G+v}=\cT_\G+\cT_v$
is also an integral of motion.  

{\bf Example 1.} Let us consider a one-dimensional motion of a point particle
subject to friction. It is described by the dynamic equation
\mar{n106}\beq
q_{tt}=-kq_t, \qquad k>0, \label{n106}
\eeq
on the configuration space $Q=\Bbb R^2\to\Bbb R$ coordinated by $(t,q)$.
This equation is equivalent to the Lagrange equation of the Lagrangian
\be
L=\frac12 \exp[kt]q^2_t dt. 
\ee
It is a hyperregular Lagrangian. The unique associated Hamiltonian form
reads
\mar{n105}\beq
H=pdq-\frac12 \exp[-kt]p^2 dt. \label{n105}
\eeq
The corresponding Hamilton equation
\be
q_t=\exp[-kt]p, \qquad p_t=0
\ee
is equivalent to the dynamic equation (\ref{n106}).
Let us consider the vector field
\be
\G=\dr_t -\frac{k}{2}q\dr_q
\ee
on $Q\to\Bbb R$.
Its prolongation (\ref{n8}) onto $V^*Q$ reads
\be
\wt\G=\dr_t -\frac{k}{2}q\dr_q +\frac{k}{2}p\dr^p.
\ee
It is readily observed that the Lie derivative of the Hamiltonian form
(\ref{n105}) with respect to this vector field vanishes. Then, the
energy function 
\mar{n107}\beq
\cT_\G=\frac12(\exp[-kt]p^2 +kqp) \label{n107}
\eeq
is an integral of motion. Another integral of motion is the momentum
$\cT_u=-p$ along the vertical vector field $u=\dr_q$.

Note that, given a connection $\G$ (\ref{n1}),
there exist bundle coordinates on 
$Q\to\Bbb R$ such that 
$\G_i=0$. A glance at the
expression (\ref{n42}) shows that the Lie derivative $\bL_{\wt\G}H$ 
vanishes iff the Hamiltonian $\cH$ written with respect
to these coordinates is independent of time. 

In Example 1, such a coordinate is 
\mar{n110}\beq
q'=\exp\left[\frac{k}{2}t\right]q. \label{n110}
\eeq
Accordingly, we have
\mar{n111}\beq
p'=\exp\left[-\frac{k}{2}t\right]p. \label{n111}
\eeq
The Hamiltonian form (\ref{n105}) with respect to these coordinates is
\mar{n112}\beq
H=p'dq'-\frac12(p'^2+q'p') dt. \label{n112}
\eeq

\section{Quantum Noether conservation laws}

In order to quantize non-relativistic time-dependent mechanics, 
we provide geometric quantization of the cotangent bundle $T^*Q$ with
respect to vertical polarization which is the vertical tangent bundle
$VT^*Q$ of $T^*Q\to Q$ (see \cite{jmp02,jmp021} for a detailed exposition).
This quantization is compatible with the Poisson algebra monomorphism
$C^\infty(V^*Q)\to C^\infty(T^*Q)$. The corresponding quantum algebra
consists of smooth functions which are affine in momenta $p$, $p^i$.
Moreover, we restrict our consideration to its subalgebra $\cA$ of functions
\mar{n120}\beq
f=a^tp +a^i(t,q^j)p_i +b(t,q^j), \qquad a^t=0,1. \label{n120}
\eeq
These functions are represented by the operators
\mar{n121}\beq
\wh f= -ia^t\dr_t -ia^k\dr_k -\frac{i}{2}\dr_ka^k + b \label{n121}
\eeq
which act in the space $E$ of sections $\psi$ of the fibre bundle $\cD\to Q$
whose restriction $\psi_t$ to each fibre $Q_t$ of $Q\to\Bbb R$ are complex
half-forms of compact support on $Q_t$. This space is
provided with the structure of a pre-Hilbert $C^\infty(\Bbb R)$-module
with respect to the non-degenerate Hermitian forms
\be
\lng \psi_t|\psi'_t\rng_t=\op\int_{Q_t}\psi_t\psi'_t.
\ee
One can use the formal coordinate expression 
\be
\psi=\psi(t,q^j)(dq^1\w\cdots\w dq^m)^{1/2}.
\ee
The operators (\ref{n121}) obey the Dirac
condition 
\be
[\wh f,\wh f']=-i\wh{\{f,f'\}_T}.
\ee
They are Hermitian because of the equality
\be
&& \lng\wh f\psi_t|\psi_t\rng-\lng\psi_t|\wh f\psi_t\rng_t=\\
&& \qquad \op\int_{Q_t}[\ol\psi(
-ia^t\dr_t -ia^k\dr_k -\frac{i}{2}\dr_ka^k + b)\psi
-\psi(ia^t\dr_t +ia^k\dr_k +\frac{i}{2}\dr_ka^k + b)\ol\psi= \\
&&\qquad -ia^t\dr_t
\lng\psi_t|\psi_t\rng_t -i\op\int_{Q_t}\dr_k(a^k\psi\ol\psi) = -ia^t\dr_t
\lng\psi_t|\psi_t\rng_t.
\ee

It should be emphasized that the homogeneous Hamiltonian $\bH$
(\ref{n70}) need not belong to the quantum algebra $\cA$, unless $\cH$
is affine in momenta $p^i$. Let us further assume that $\bH$ is a
polynomial of momenta. One can show that, in this case,
$\bH$ can be represented by a product of
affine functions (\ref{n120}) and, consequently, can be quantized as a
Hermitian element 
\mar{n122}\beq
\wh\bH=-i\dr_t +\wh\cH \label{n122}
\eeq
of the 
enveloping algebra of the Lie algebra $\cA$ \cite{jmp02,jmp021}.
However, this representation and the corresponding quantization
(\ref{n122}) fail to
be unique. The operator (\ref{n122}) yields the Schr\"odinger
equation
\mar{n123}\beq
\wh\bH\psi=(-i\dr_t +\wh\cH)\psi=0, \qquad \psi\in E. \label{n123}
\eeq

Given quantizations (\ref{n121}) and (\ref{n122}), any Noether symmetry
function $\cT_u$ (\ref{n50}) is quantized by the Hermitian operator
\mar{n124}\beq
\wh\cT_u=u^t\wh\cH-\wh{u^ip_i} \label{n124}
\eeq
in the pre-Hilbert module $E$. If $\cT_u$ is an integral of motion, the
operator $\wh\cT_u$ (\ref{n124}) commutes with the Schr\"odinger
operator $\wh\bH$, 
and acts in the subspace $\Ker\wh\bH\subset E$ of solutions of the
Schr\"odinger equation (\ref{n123}).

In particular, let $u=v=v^k\dr_k$ and $\cT_v=-v^k(t,q^j)p_k$. Since the
operator  $\wh\cT_v=iv^k(t,q^j)\dr_k$ commutes
with the Schr\"odinger
operator $\wh\bH$ (\ref{n123}), we obtain the equalities 
\be
&& \dr_t \lng \wh\cT_v\psi|\psi\rng=\op\int_{Q_t}[(i\dr_tv^k\dr_k\psi)
\ol\psi +(iv^k\dr_k\dr_t\psi)\ol\psi +(iv^k\dr_k\psi)\dr_t\ol\psi]=\\
&& \qquad \op\int_{Q_t}[(i\dr_tv^k\dr_k\psi)
\ol\psi +v^k\dr_k(\cH\psi)\ol\psi -(v^k\dr_k\psi)\ol{(\cH\psi)}]=
-\op\int_{Q_t}([\wh\bH,\wh\cT_v]\psi)\ol\psi=0
\ee
on the $\Ker\wh\bH$. It follows that the mean values of the operator
$\wh\cT_v=iv^k\dr_k$ on solutions of the Schr\"odinger equation
(\ref{n123}) are conserved.

Let $u=\G$ and $\cT_\G=\cH-p_k\G^k$. Let us choose bundle
coordinates $q'^j$ such that $\G^k=0$. Then, $\cT_\G=\cH$ is independent
of time. Since $\wh\bH=-i\dr_t+\wh\cT_\G$, the Schr\"odinger equation
(\ref{n123}) comes to the conservative one
\mar{n130}\beq
\psi=\exp(-i\cE t)\f_\cE, \qquad \wh\cT_\G\f_\cE=\cE\f_\cE. \label{n130}
\eeq

For instance, the Schr\"odinger equation for a point particle subject
to friction in Example 1 reads 
\be
-\frac12(\dr_{q'}^2+iq'\dr_{q'})\f_\cE=\cE\f_\cE.
\ee

\end{document}